\documentclass[sigconf]{acmart}

\usepackage[utf8]{inputenc}
\usepackage{paralist}
\usepackage{algorithm}
\usepackage[noend]{algpseudocode}
\usepackage{xcolor}
\usepackage{verbatim}
\usepackage{multirow}
\usepackage{listings}
\newcommand{\toolname}{\texttt{RTj}}
\newcommand{\testanalyzer}{\texttt{Test Analyzer}}

\definecolor{pblue}{rgb}{0.13,0.13,1}
\definecolor{pgreen}{rgb}{0,0.5,0}
\definecolor{pred}{rgb}{0.9,0,0}
\definecolor{pgrey}{rgb}{0.46,0.45,0.48}
\definecolor{light-gray}{gray}{0.80}
\newcommand{\dest}[1]{\emph{{#1}}}

\title{\toolname: a Java framework for detecting  and refactoring rotten green test cases}

\begin{document}

\author{Matias Martinez}
\email{matias.martinez@uphf.fr}
\affiliation{%
\institution{Universit\'e Polytechnique Hauts-de-France, LAMIH UMR CNRS 8201, France}
}

\author{Anne Etien}
\email{anne.etien@univ-lille.fr}
\affiliation{%
  \institution{Universit\'e de Lille, CNRS, Inria, Centrale Lille, UMR 9189 – CRIStAL, France}
}

\author{St\'ephane Ducasse}
\email{stephane.ducasse@inria.fr}
\affiliation{%
  \institution{Inria, Universit\'e de Lille, CNRS, Centrale Lille, UMR 9189 – CRIStAL, France}
}
\author{Christopher Fuhrman}
\email{christopher.fuhrman@etsmtl.ca}
\affiliation{%
  \institution{\'Ecole de technologie sup\'erieure, Montr\'eal, Qu\'ebec, Canada}
}

\begin{abstract}
Rotten green tests are  passing tests which have, at least, one assertion not executed.
They give  developers a false confidence.
In this paper, we present, \toolname{}, a framework that analyzes test cases from Java projects with the goal of  detecting and refactoring rotten test cases. 
\toolname{} automatically discovered 427 rotten tests from 26 open-source Java projects hosted on GitHub.
Using \toolname{}, developers have an automated recommendation of the tests that need to be modified for improving the quality of the applications under test.
\end{abstract}

\maketitle

\thispagestyle{plain}
\pagestyle{plain}

\section{Introduction}

Software developers write unit test cases with the goal of improving code quality and preventing code regression.
Passing (green) tests are usually taken as a robust sign that the code under test is valid \cite{Delplanque:2019:RGT}.
However, a passing test can have, at the same time, a poor design, which detracts from maintainability.
Such cases are known as \emph{Smelly Tests} \cite{Deursen:2001:RTC,Reic07a}.

A \emph{Rotten Green Test} is even a stronger test problem: a rotten green test is a test that passes (is green) but contains assertions that are never executed \cite{Delplanque:2019:RGT}. 
Rotten tests give developers false confidence because, beyond passing, an assertion that should validate some property is, in fact, not executed.
Previous work has shown the presence of rotten test cases in software written in Pharo \cite{Delplanque:2019:RGT}.

In this paper, we present \toolname, a framework for detecting and refactoring smelly tests written in Java. 
The current implementation analyzes JUnit tests and classifies them according to the categories of rotten green tests presented in \cite{Delplanque:2019:RGT}.
For doing that analysis, \toolname{} takes as input the source code of the program under analysis including their test cases. It does a static analysis for detecting the code elements from a test case (assertion, helpers, etc.) and a dynamic analysis to determine the execution of such elements.
\toolname{} produces as output a detailed report with all rotten and smelly tests found and, when possible, a refactored version of such tests.

We executed \toolname{} on 67 open-source projects written in Java hosted on GitHub with more than 1000 stars and 100 forks. 
We found 427 rotten test cases from 26 projects.
Our results show the importance of having a tool that analyzes the quality of test cases:
developers from one third of the projects analyzed trust the passing tests which, beyond having assertions that validate some properties, are not executed.
\toolname{} can be used by both researchers and software practitioners.
\emph{Researchers} can use it for carrying out different empirical studies of test cases, and to write, using the extension mechanism proposed by \toolname, analyzers able to detect new cases of rotten and smelly tests.
\emph{Software practitioners} can use \toolname{} for analyzing and refactoring their own software with the goal of improving the quality of their applications.

This paper continues as follows:
Section \ref{sec:testcharacterization} defines rotten green tests and its categories.
Section \ref{sec:arch} presents the architecture of \toolname.
Section \ref{sec:evaluation} presents  rotten tests found by \toolname{}.
Section \ref{sec:rw} presents the related work.
Section \ref{sec:discussion} presents a discussion about the future work around \toolname.
Section \ref{sec:conclusion} concludes the paper.

\toolname{} is publicly available at: \url{https://github.com/UPHF/RTj}

\section{Test Case Categorization}
\label{sec:testcharacterization}
A rotten green test contains a call site for an assertion primitive or a test helper but this assertion or helper was not invoked during test execution. Not all rotten green tests are caused by the same problem. This section gives a brief description of the four categories presented in \cite{Delplanque:2019:RGT}.

\begin{itemize}
\item \dest{A Context-dependent test} contains conditionals with different assertions in the different branches. 
\item \dest{A Missed Fail test} contains an assertion which was forced to fail. 
\item \dest{A Skip test}  contains guards to stop their execution early under certain conditions.
\item \dest{Fully Rotten tests} do not execute one or many assertions, and do not fall into any of the previous categories.
\end{itemize}

\begin{algorithm}[t]
\label{algo:main1}
\hspace*{\algorithmicindent} \textbf{Input: program under analysis $P$} \\
\hspace*{\algorithmicindent} \textbf{Output: Labels and refactored tests} 

\begin{algorithmic}[1]

\State $M \gets createAppModel(P)$ \label{algo:createmodel}
\State  $T \gets findTestCases(P, M)$\label{algo:findtest}
\State  $Dit \gets executeTest(P, T)$\label{algo:executetest}

\State  $labels \gets \emptyset, refactors \gets \emptyset$, \State $stat\_results \gets \emptyset$, $dyn\_results \gets \emptyset$

\For{$t_{i}$ in $T$}\label{algo:fortest}

\For{$a_{i}$ in $Test Analyzers$} \label{algo:foranalstat}
    \State $s_i = a_{i}.findElements(M, stat\_results, t_i)$  \label{algo:findelements} %
     \State $stat\_results = stat\_results \; \cup <a_i, s_i>$   \label{algo:savestatic}
   \State $d_i = a_{i}.dynamicAnalysis(M, Dit, dyn\_results, s_i, t_i)$\label{algo:dynanal} %
   \State $dyn\_results = dyn\_results \; \cup <a_i, d_i>$  \label{algo:savedyn}
    \State $l_i = a_{i}.labelTest(M, s_i, d_{i},  t_i)$ \label{algo:label}
     \State $labels = labels \; \cup <t_i, l_i>$ \label{algo:savelabel}
     \State $t_i' = a_{i}.applyRefactor(M, s_i, d_{i},  t_i)$  \label{algo:refactor}
     \If{$t_i' \; is \; not \; null$}  \label{algo:saverefactor}
             \State $refactors = refactors \; \cup <t_i, t_i'>$ \label{algo:}
      \EndIf
\EndFor

\EndFor  
\Return $labels, refactors$

\end{algorithmic}
\caption{\toolname: analysis and refactor of test cases.}
\label{alg:the_alg}
\end{algorithm}

\section{Architecture}
\label{sec:arch}

\toolname{} executes the  steps presented in Algorithm \ref{alg:the_alg}.
This section describes each of them.
The input is the source code of  program under analysis, including its test cases ($P$).
The output is twofold:
\begin{inparaenum}
\item a list of the test cases analyzed with labels, and 
\item refactored tests.
\end{inparaenum}

\subsection{Basic analysis steps}
\paragraph{Step 1: Creation of the program model}
\label{sec:modelcreation}

First, \toolname{} creates a model that represents $P$ and its test cases (line \ref{algo:createmodel}).
\toolname{} takes as input the source code of $P$ and generates a model based on Spoon meta-model \cite{spoon}. The generated model is an enriched abstract syntax tree (AST).

\paragraph{Step 2: Test cases detection}
\label{sec:tcdetection}
\toolname{} searches for all test cases written in $P$ (line \ref{algo:findtest}).
For that, \toolname{} filters from the model all code elements (e.g., methods) that correspond to test cases.
By default, \toolname{} analyzes projects that use the testing framework 4.X. Consequently, one of the heuristics \toolname{}  applies is to filter methods with the annotation \emph{org.junit.Test}.

\paragraph{Step 3: Instrumented test case execution}
\label{sec:testexecution}

\toolname{} executes the test cases in an instrumented version of $P$ (line \ref{algo:executetest}).
The goal of the instrumentation is to trace the executed lines (from both the application and tests) by each test case. 
The output  of this step  ($Dit$ on line \ref{algo:executetest}) is twofold:
\begin{inparaenum}[\it 1)]
\item the result of each test case (e.g., passing, failing),
\item for each line $l$ from $P$, it contains the test cases that executed $l$ and the number of times that $l$ was executed.
Note that if even a line $s$ belongs to a test, it can be executed zero times  (e.g., $s$ is inside an \texttt{If} statement).
\end{inparaenum}

\subsection{Test analysis}
\label{sec:testanalyzer}
\toolname{} iterates over the list of test cases found (line \ref{algo:fortest}) to analyze each of them ($t_i$) using \testanalyzer s.

A  \testanalyzer{} is a component that takes as input a test $t_i$, analyzes it given a specific goal and, possibly,  proposes $t'_i$, a refactored version of $t_i$.
\toolname{} has at least one \testanalyzer{} for each rotten category presented in Section \ref{sec:testcharacterization}.
For example, for \emph{Context Dependent} there are one analyzer for detecting rotten assertion from a test, another for detecting rotten calls to helper, and a third one for detecting rotten assertions inside an invoked helper method.
\toolname{} iterates over the \testanalyzer{} (line \ref{algo:foranalstat}), and applies them on each test $t_i$.

A \testanalyzer{} has four main responsibilities, which are described below.

\emph{1) Identification of test elements:}
\label{sec:identifytestelement} a \testanalyzer{} does a static analysis (line \ref{algo:findelements}) which consists of parsing the generated model to identify all code elements (e.g., Assertions, Helpers, fails, returns) that it needs to classify a test into a specific category. 

\emph{2) Dynamic analysis of test elements:}
A \testanalyzer{} analyzes the execution of the elements it filtered in the previous step (line \ref{algo:dynanal}). 
\toolname{} provides a procedure which determines, given the dynamic analysis information ($Dit$), whether an element was executed or not during the execution of a test case $t_i$.

\emph{3) Test classification:} 
\label{sec:tclassification}
A \testanalyzer{} classifies a  test case $t_i$ (line \ref{algo:label}).
It receives as parameters the elements that are useful for its goal, the dynamic information for them (i.e., if those were executed by $t_i$), and produces as output a set of \emph{labels} (possibly empty), where each label indicates a particular classification of $t_i$ such as \emph{Fully Rotten test}, \emph{Missed fail},  etc.

\emph{4) Test Refactor:}
\label{sec:testrefactor}
A \testanalyzer{} is able to propose to the user the candidate refactorings of the test case $t_i$ (line \ref{algo:refactor}).
It receives as input the generated model, and creates the refactoring by transforming a cloned model.
The Spoon model created by default has an API for applying transformations (e.g., to replace, insert or remove a code element) and generates source code from a modified model.

\vspace{-0.4cm}
\subsection{\testanalyzer s  included in \toolname}
\toolname{} implements \testanalyzer s that are able to detect all categories of rotten green tests defined in \cite{Delplanque:2019:RGT} and described in Section \ref{sec:testcharacterization}.
We now briefly describe some of them.

\emph{Assertion Rotten analyzer:} It determines if a test has rotten assertions by parsing the test's model to identify static method invocations whose names start with the keyword ``assert" and the class of the invocation's target is \texttt{org.junit.Assert}.
Then, this analyser labels a test $t_i$ as `Assertion Rotten' if one of those assertions was not executed. 
\toolname{} splits this analyzer in two:
\begin{inparaenum}
\item \emph{Context-dependent assertion rotten test}: 
if the rotten element is inside an \texttt{if-else}.
\item \emph{Fully rotten assertion test}: if it is inside another element.
\end{inparaenum}

\emph{Rotten Call Helper analyzer:} Unlike the previous analyzer, it looks for method calls to helpers.
It determines that a method $e$ is a helper if it has:
\begin{inparaenum}
\item  an assertion, or
\item an invocation to a method helper.
\end{inparaenum}
\toolname{} also distinguishes between 
\begin{inparaenum} 
\item \emph{Context-dependent}, and
\item \emph{Fully rotten} cases.
\end{inparaenum}

\emph{Rotten Assertion in Helper analyzer:} This analyzer checks whether an invoked helper $h$ does not execute an assertion written in $h$.

\emph{Skip Test analyzer:} identifies \texttt{return} statements in test $t_i$.
Then, it classifies a  $t_i$ as ``Skip'' if the \texttt{return} was executed and, there are no executed assertions written  below that return.

\emph{Missed Fail analyzer}: searches for assertions that are forced to fail, e.g., \texttt{assertTrue(false).}
This analyzer does not check if such assertion were executed.

\emph{Smoke test}: neither contains any assertion nor helper calls. 
Note that \cite{Delplanque:2019:RGT} does not categorize a smoke test as rotten green test.

The current implementation of \toolname{} provides {2} refactorings.
\emph{Replacement of missing fail}: the missing fail analyzer proposes a refactoring that replaces the assertions forced to fail by invocations to $org.junit.Assertion.fail()$.

\emph{Add  comment}: all analyzers can add a \texttt{TODO} comment just before the rotten code element.
IDEs such as Eclipse display such \texttt{TODO} comments in a dedicated view.

\section{Evaluation}
\label{sec:evaluation}

\subsection{Methodology}
\label{sec:methodology}
We aimed to study \emph{popular} open-source Java projects.
For that, we first selected from GitHub projects having:
\begin{inparaenum}
\item Java as main language,
\item JUnit 4 as testing framework,
\item Maven as dependency manager system,
\item more than 1000 stars and 100 forks.
\end{inparaenum}
Then, we executed \toolname{} on 67 of them.
The execution time of \toolname{} mostly depends on the execution time of the instrumented test cases and on the number of test cases. It took from 1 minute (Streamex project) to 20 minutes (XChange).

\subsection{Results}

Table \ref{tab:rotten:results} summarizes the results and shows the number of rotten test cases (by category) from the 10 projects with larger number of rotten tests.

\toolname{} found in total 427 rotten green tests on 26 projects.
This means that the 38\% of the project analyzed (26 out of 67)  have, at least, one rotten green test.
The majority of the rotten tests found by \toolname{}  are from two rotten categories: 253 Context dependent tests from 18 projects, and 110 Fully rotten tests from 16 projects (in total 23 distinct projects).
This means that around one out of three projects (23/67) has passing tests that do not execute assertions written in such tests.
Developers of such projects are trusting the results of those rotten tests (\emph{passing}); however, they are very likely unaware that some validations written in them are not being executed.

\begin{table}[t!]
 \caption{
 The table shows the 10 projects with largest number of rotten green tests. At the bottom, it summarizes the total of rotten tests found and the number of projects affected by each rotten category.}
 \vspace{-0.4cm}
\centering
 \begin{tabular}{|l | r r r r | r|} 
 \hline
Top-10&Context & Missed  & Skip  &Fully &\multirow{2}{*}{Total} \\ 
projects& dependent&  Fail &  Test & Rotten &\\ 
\hline
\hline
Optaplanner&	104	&	0	&	0	&	7	&	111	\\
Flink-core&	26	&	2	&	9	&	10	&	47	\\
Streamex&	39	&	0	&	0	&	2	&	41	\\
Bt&	33	&	0	&	1	&	4	&	38	\\
XChange&	1	&	0	&	15	&	22	&	38	\\
Handlebars&	0	&	0	&	0	&	30	&	30	\\
Joda-time&	5	&	3	&	17	&	1	&	26	\\
Jeromq&	12	&	0	&	0	&	4	&	16	\\
Wasabi&	0	&	0	&	0	&	15	&	15	\\
Mahout&	1	&	0	&	5	&	1	&	7	\\

\hline
\hline
Total & \multicolumn{5}{|c|}{}\\
\hline
Rotten Tests &253	&	16	&	48	&	110	&	427\\
Projects affected&18	&	6	&	6	&	16	&	26\\

 \hline
 \end{tabular}

 \label{tab:rotten:results}
\end{table}

\subsection{Illustrative cases}
\label{sec:results}
This section presents one rotten green tests found by \toolname{} test per rotten category.

\subsubsection{Context Dependent Rotten Assertion Test}
\label{sec:cdtest}
Test {LambdaExtractionTest}. testCoGroupLambda() from project Apache-Flink belongs to this category. %

\lstset{
  numbers=left,
  firstnumber=206,
  numberfirstline=true,
  tabsize=1
}
\begin{lstlisting}[escapechar=~, basicstyle=\tiny, label=lst:contex1, belowskip=-1pt, aboveskip=-0pt]
@Test public void testCoGroupLambda() {
		CoGroupFunction<Tuple2<...>> f = (i1, i2, o) -> {};
		TypeInformation<?> ti = TypeExtractor.getCoGroupReturnTypes(f, ...);
		if (!(ti instanceof MissingTypeInfo)) {
            ~\colorbox{light-gray}{assertTrue(ti.isTupleType());}~
            ~\colorbox{light-gray}{assertEquals(2, ti.getArity());}~ 
        ...	}
	}
\end{lstlisting}

The test has rotten assertions located inside the \texttt{Then} branch of an \texttt{If} (line 209), whose condition's evaluation is always \texttt{false}.

\subsubsection{Fully Rotten Test}

Test {BucketLeapArrayTest}.   {testListWindowsNewBucket()} from project Alibaba-Sentinel is Fully rotten.

\lstset{
  numbers=left,  
  firstnumber=209,
  numberfirstline=true,
  tabsize=1,
}
\begin{lstlisting}[escapechar=~, basicstyle=\tiny, label=lst:fullyrotten1, belowskip=-1pt,  aboveskip=-0pt]
    @Test
    public void testListWindowsNewBucket() throws Exception {
        ...
        BucketLeapArray leapArray = new BucketLeapArray(sampleCount, intervalInMs);
        ....
        List<WindowWrap<MetricBucket>> list = leapArray.list();
        for (WindowWrap<MetricBucket> wrap : list) {
            ~\colorbox{light-gray}{assertTrue(windowWraps.contains(wrap));}~
        }
\end{lstlisting}

The test has a \texttt{For} that iterates over a list. Inside the loop body, the test has an assert (line 216). 
As the list is always empty, the assertion is never executed.

\subsubsection{Skip Rotten Test}
\label{sec:skiptest}
Test {testNormalizedKeyReadWriter()} written in test helper {ComparatorTestBase} from project Apache-flink is an Skip test.

\lstset{
  numbers=left, 
  firstnumber=371,
  numberfirstline=true,
  tabsize=1,
}
\begin{lstlisting}[escapechar=~, basicstyle=\tiny, label=lst:skip1, belowskip=-1pt,  aboveskip=-0pt]
@Test public void testNormalizedKeyReadWriter() {
		...
			TypeComparator<T> comp1 = getComparator(true);
			if(!comp1.supportsSerializationWithKeyNormalization()){
					~\colorbox{light-gray}{return}~;
			}
			...
			~\colorbox{light-gray}{assertTrue(comp1.compareToReference(comp2) == 0);}~
		...
	}
\end{lstlisting}

Every execution of this helper does not execute any assertion written in it (e.g., line 378), because the guard at line 374 is always \texttt{true}. Thus the \texttt{return} from line 375 is always executed.

\subsubsection{Missed fail}
\toolname{} also detected instances of missed fail.
For example, in test testHasProtectedConstructor from the Reflectasm project,
the developer used \texttt{assertTrue(false)} instead of using \texttt{fails()}.

\subsection{False Positive Cases}
\label{sec:fpcontexttest}

As reported by \cite{Delplanque:2019:RGT}, the detection of rotten test cases in Pharo suffered the presence of false positives due to conditional use or multiple test contexts.
We have implemented some heuristics in \toolname{} to detect such cases and present them as special cases.
For instance, \toolname{} labels a test $t$ as ``Both-branches-with-Assertion" Context-dependent when $t$:
\begin{inparaenum}[\it 1)]
\item has an \texttt{If} with \texttt{Then} and \texttt{Else} branches, 
\item both branches have an assertion or a helper call, and
\item only one branch is executed.
\end{inparaenum}
One such case is 
test testJdk9Basics() from project Streamex, 
a library for enhancing Java 8 Streams.

\lstset{
  numbers=left,  
  firstnumber=59,
  numberfirstline=true,
  tabsize=1,
}
\begin{lstlisting}[escapechar=~, basicstyle=\tiny, floatplacement=tbp,belowskip=-1pt,aboveskip=1pt]
@Test public void testJdk9Basics() {
        MethodHandle[][] jdk9Methods = Java9Specific.initJdk9Methods();
        if (Stream.of(Stream.class.getMethods()).anyMatch(m -> m.getName().equals("takeWhile")))
            assertNotNull(jdk9Methods);
        else
            assertNull(jdk9Methods);
    }
    \end{lstlisting}

The test has an \texttt{If} with assertions in the two branches: one for asserting Java 9 code, the other for asserting the others versions.
The test executes only one single branch and it depends on the JDK used for running it.

\section{Related Work}
\label{sec:rw}

ReAssert \cite{Daniel:2011:RTR} and TestCareAssistant \cite{Mirzaaghaei:2010:ART}, are two tools that can automatically suggest repairs for broken Junit tests.
Our tool focuses on analyzing and refactoring passing (i.e., no broken) tests.

Deursen et al. \cite{Deursen:2001:RTC} presented 11  bad code smells that are specific for test code and proposed 6 test refactorings that aim to improve test understandability, readability, and maintainability. Reichnard et al. \cite{Reic07a} presented a tool and an extended list of identified test smells.
Oliveto et al. \cite{Oliveto:2012:EAD} have conducted an empirical study for 
analyzing the distribution of 9 test smells from \cite{Deursen:2001:RTC} in real software applications. They found that the 82\% of JUnit classes analyzed were affected by at least one test smell.
This result shows the importance of providing developers a unified framework for detecting and refactoring smelly tests.

\section{Discussion}
\label{sec:discussion}
\subsection{Extending \toolname}
\label{sec:extensionpoint}
\toolname{} provides \emph{Extension points} that allow users to override the default behaviour of the framework and to add new functionality. 

\emph{Model Creation:}
\label{sec:extpoint:modelcreation}
\toolname{} provides an extension point to override the procedure that creates the model of the program under analysis (Section \ref{sec:modelcreation}).
This allows \toolname{} to use other program meta-models. 
For instance, an extension could create a FAMIX Java model \cite{ducasse:hal-00646884} using the tool VerveineJ.\footnote{\url{https://github.com/moosetechnology/VerveineJ}}
Note that the use of a new meta-model $M$ requires that the \emph{Test Analyzers} be capable of analyzing  and transforming the new model built from $M$.

\emph{Execution of test cases:}
\label{sec:extpoint:testcaseexect}
By default, \toolname{} is able to run JUnit 4.X and to detect element defined in JUnit's API (e.g. Assertion).
This extension point makes it possible to execute other testing frameworks such as JUnit 5 or TestNG.

\emph{Addition of new Test Analyzer:}
\label{sec:extpoint:analyzers}
This extension point allows users to plug new \emph{Test Analyzers} for studying cases not already considered.
\toolname{} represents a \testanalyzer{} as a Java interface that declares four methods, and each one corresponds to one step presented in section \ref{sec:testanalyzer}.
By default, analyzers must interact with a Spoon model representing the program under analysis. 
The Spoon meta-model was designed to be easily understandable by Java developers, and provides developers different well-documented mechanisms to write program analyses and transformations.

\emph{Result output:}
\label{sec:extpoint:output}
By default, \toolname{} generates a Json file that lists all the cases found by the analyzers.
\toolname{} provides an extension point for generating new types of outputs (e.g., reports).

\subsection{Future work}
\label{sec:fw}

We are currently performing a large-scale empirical study on rotten green tests from popular open-source Java projects (in terms of stars, forks, commits, committers, etc) hosted on GitHub. 

Beyond rotten green tests, we will focus on other kinds of smelly tests, and even on refactorings that have other objectives.
For instance, we are currently developing an analyzer in \toolname{} that detects and refactors \emph{Impure test} \cite{Xuan:2016:BRE}, with the goal of improving dynamic analysis tasks such as fault localization and automated program repair.

We will add the ability to analyze tests from other Java testing frameworks such as JUnit 5 and TestNG, and propose refactors for the rotten green tests presented in this paper.
Moreover, as the current implementation of \toolname{} provides a command-line interface, we plan to create plug-ins for Maven and  IDEs (Eclipse and IntelliJ).

\section{Conclusion}
\label{sec:conclusion}
This paper presents \toolname{}, a framework that analyzes (statically and dynamically) test cases with the goal of detecting smelly tests, including rotten green test.
Rotten green test is a serious problem because give developers false confidence in the system under tests.
\toolname{} aims at helping developers to automatically detect those smelly tests, which can then be refactored to improve the quality of an application.
\toolname{} also proposes to developers candidate test refactorings.
Using \toolname{} we found 427 rotten green tests from 26 open-source projects hosted on GitHub.
The design of \toolname{} allows users to extend the framework by adding new analyzers focusing on other kinds of smelly tests and on other testing frameworks.

\toolname{} is publicy available at: \url{https://github.com/UPHF/RTj}

\bibliographystyle{plain}
\bibliography{references}

\end{document}